\documentclass[aps,pra,twocolumn,noshowpacs,superscriptaddress,groupedaddress]{revtex4}  
\usepackage{graphicx}  
\usepackage{dcolumn}   
\usepackage{bm}        
\usepackage{amssymb}   
\usepackage{amsmath}
\usepackage{mathtools}
\usepackage{color}
\usepackage[free-standing-units=true]{siunitx} 
\usepackage[colorlinks=true, pdfstartview=FitV, linkcolor=blue, citecolor=blue, urlcolor=blue]{hyperref} 
\usepackage{epstopdf}

\usepackage{appendix} 

\newcommand{\affilITP}{Institute for Theoretical Physics, ETH Z\"{u}rich, CH-8093 Z\"urich, Switzerland.}
\newcommand{\affilIFP}{Laboratory for Solid State Physics, ETH Z\"{u}rich, CH-8093 Z\"urich, Switzerland.}

\begin{document}

\preprint{}

\title{Ising machines with strong bilinear coupling}

\author{Toni L. Heugel}\affiliation{\affilITP}
\author{Oded Zilberberg}\affiliation{\affilITP}
\author{Christian Marty}\affiliation{\affilIFP}
\author{R. Chitra}\affiliation{\affilITP}
\author{Alexander Eichler}\affiliation{\affilIFP}
\date{\today}

\begin{abstract}

Networks of coupled parametric resonators (parametrons) hold promise for parallel computing architectures.
En route to realizing complex networks, we report an experimental and theoretical analysis of two coupled parametrons. In contrast to previous studies, we explore the case of strong bilinear coupling between the parametrons, as well as the role of detuning. We show that the system can still operate as an Ising machine in this regime, even though careful calibration is necessary to ensure that the correct solution space is available. Apart from the formation of split normal modes, new states of mixed symmetry are generated. Furthermore, we predict that systems with $N>2$ parametrons will undergo multiple phase transitions before arriving at a regime that can be equivalent to the Ising problem.
\end{abstract}
\maketitle

\section{Introduction}
\begin{figure}[t!]
  \includegraphics[width=\columnwidth]{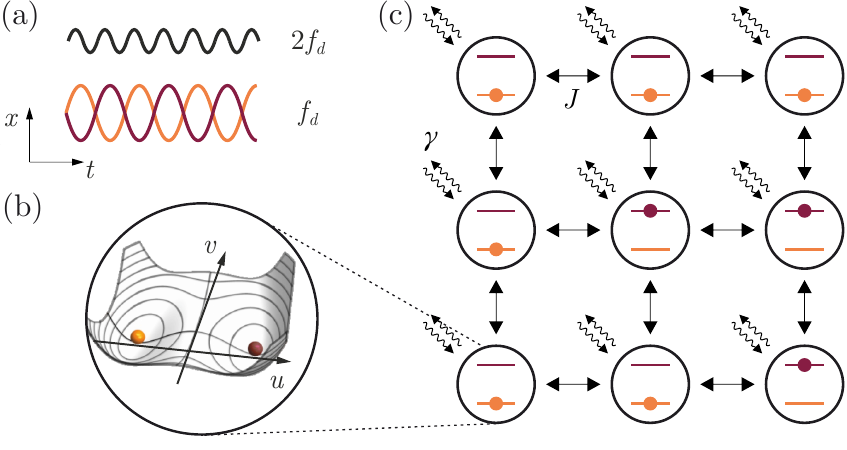}
  \caption{Parametron networks. (a)~In response to a parametric drive at frequency $2f_d$ (black), two possible stable oscillation states with a phase difference of $\pi$ (wine red, orange) emerge at subharmonic frequency $f_d \approx f_0$; $x$ is displacement and $t$ is time. (b)~Rotating-frame energy landscape of the parametric resonator in a phase space rotating at $f_d$, with $x = u \cos(2\pi f_d t) - v \sin(2\pi f_d t)$. The stable phase states are indicated by spheres. (c)~Schematic representation of a network of parametrons, with a coupling rate $\gamma$ to a bath and nearest-neighbor coupling rate $J$. Each parametron acts as a bistable element.
  }
  \label{fig:Fig1}
\end{figure}

Driven nonlinear systems were first considered as logic elements at the dawn of the digital era~\cite{Goto_1959, Neumann_1959, Sterzer_1959}. Their nonlinearity induces several stable oscillation states that can be used as elementary information units for computation and data storage. A prominent example of a nonlinear system is the parametric resonator, also known as the `Kerr nonlinear resonator' or `parametron'~\cite{Goto_1959, Neumann_1959, Hosoya_1991, Mahboob_2008}. In a space spanned by the driving amplitude $\lambda$ and driving frequency $f_d$, the parametron exhibits a well-defined `instability lobe' around the eigenfrequency $f_d \approx f_0$ and above the threshold value $\lambda_\mathrm{th}$. When driven inside this lobe, the parametron locks onto one of two `phase states' that have the same oscillation amplitude but differ by $\pi$ in phase~\cite{Landau_Lifshitz, Lifshitz_Cross, DykmanBook}, cf. Fig.~\ref{fig:Fig1}(a) and (b). These phase states represent the binary information unit of the device. Parametron-based logic operations is experiencing a resurgence of interest due to the recent development of nanomechanical, electrical and optical resonators that offer long-lived, error-resilient and tunable phase logic states~\cite{Mahboob_2008, Wilson_2010, Eichler_2011_NL, Gieseler_2012, Lin_2014, Leuch_2016, Puri_2017, Nosan_2019, Frimmer_2019, Grimm_2019, Puri_2019_PRX, Miller_2019}.

Several research fields are currently racing towards physical implementations of parametron networks for parallel computing, and their corresponding operation protocols~\cite{Gottesman_2001, Devoret_2013, Mahboob_2016, Inagaki_2016, Goto_2016, Puri_2019_PRX,Bello_2019,Okawachi_2020}. 
In a parallel network, a given task is encoded in the coupling between nodes. Under the influence of a parametric drive and the node coupling, the entire system evolves towards a stable `optimal' configuration, i.e., a particular oscillation mode involving all resonators. This oscillation mode represents the computational output, cf. Fig.~\ref{fig:Fig1}(c)~\cite{Hopfield_1982, Neural_Networks, Csaba_2016, Albash_2018, Preskill_2018}. This behavior can be exploited to solve many optimization problems that are nearly intractable with sequential computers. Examples include the famous travelling salesman problem~\cite{Lucas_2014}, the number partitioning problem~\cite{Nigg_2017}, and the MAX-CUT problem~\cite{Inagaki_2016_Science, Goto_2019}, but also fundamental questions in physics such as the ground state of the Ising spin model~\cite{Ising_1925, Rota_2019, Heim_2015} or the structure of complex molecules~\cite{Reiher_2017}.

Recently, parametron networks were explored in `Coherent Ising Machines' with dissipative coupling, corresponding to simultaneous mutual feedback between the resonators~\cite{Wang_2013,Inagaki_2016_Science,Mahboob_2016,Yamamoto_2017,Yamamura_2017,Bello_2019}. The feedback allows some parametric oscillation modes of the system to exist at a lower driving threshold $\lambda_\mathrm{th}$ than others. The network is composed of identical resonators and is operated by slowly ramping up a global parametric drive amplitude at the driving frequency $f_d = f_0$. The mode that profits from maximum positive feedback will appear at the lowest drive power. 
Note, however, that the efficiency of this Coherent Ising Machine as a way to solve computational tasks is under scrutiny~\cite{Strinanti_2021}.

Alternative types of Ising simulators are based on bilinear coupling, that is, conservative energy exchange between resonators~\cite{Goto_2016,Puri_2017_NC,Nigg_2017,Goto_2018,Dykman_2018,Rota_2019}. Previous proposals are based on the assumption of very weak coupling, such that the individual resonator states are barely affected by the interaction with other resonators.
At the same time, the proposals also rely on the notion of (quantum) adiabatic state evolution, which implies that the damping rate $\gamma$ is the smallest energy scale in the system; in particular, this requires that the coupling dominates over the dissipation. This condition, which is commonly referred to as `strong coupling', ensures that the exchange of energy (and thus information) between individual resonators is faster than the loss-induced decoherence of the network. Strong coupling, however, is expected to impact the phase diagram of the network; for instance, the available oscillation configurations (the computational solutions) can depend critically on the selected driving frequency. The behavior of a parametron network in the strong coupling regime has never been explored, and its equivalence to an Ising machine remains speculative.

In this work, we establish the validity of the adiabatic bifurcation simulator for the Ising problem in a system of two nearly identical (classical) parametrons with strong bilinear coupling.
The coupled system forms normal modes that are split in frequency, such that their corresponding instability lobes no longer have their `tips' at the same frequency~\cite{DykmanBook,Heugel_2019_TC}. In addition, the interplay between the coupling and nonlinearities leads to unexpected `mixed-symmetry' states and to hidden bifurcations that erase certain network solutions in part of the phase diagram. Nevertheless, with proper calibration, we find that the two-parametron system can be used as an Ising machine. For $N>2$ parametrons, the normal-mode perspective allows us to predict a surprising problem, arising in all instances of Ising machines: the number of available network solutions does not scale with $2^N$ close to threshold. We discuss this `state space' problem along with a potential explanation to reconcile our findings with the operational functionality of an Ising machine.

\section{Results}

\begin{figure*}[t!]
  \includegraphics[width=\textwidth]{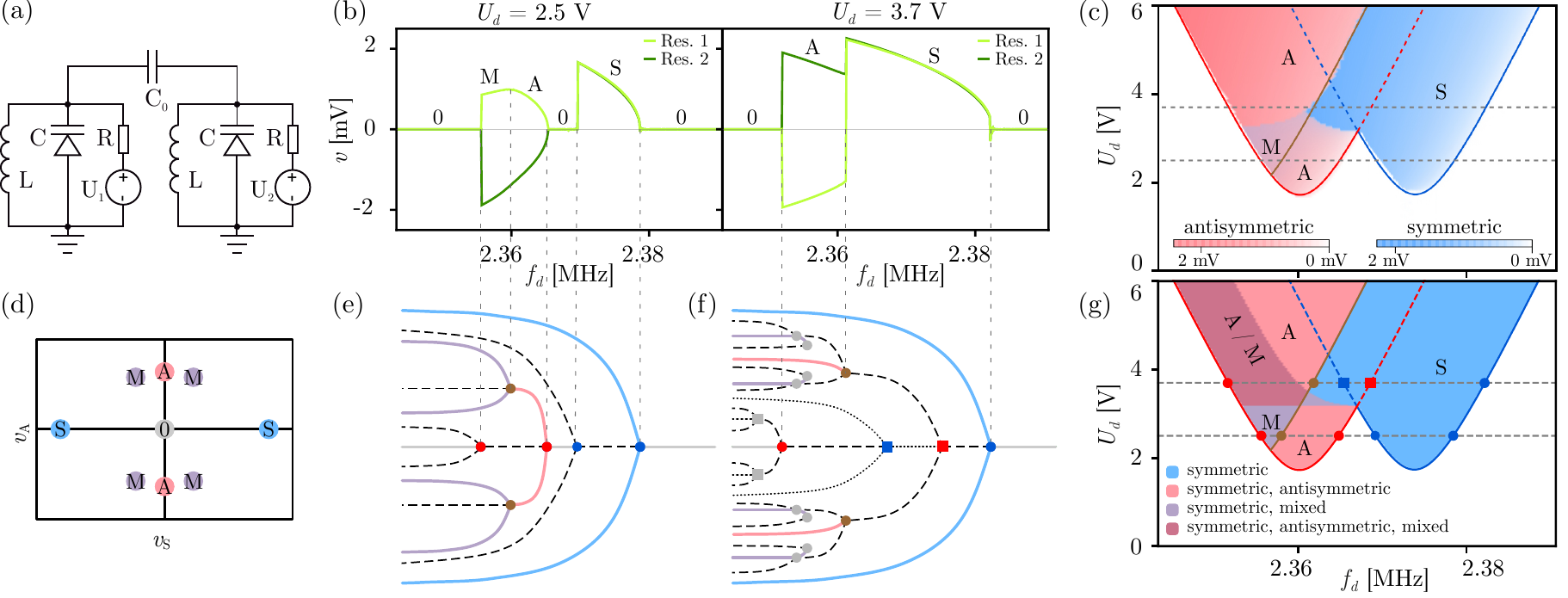}
  \caption{(a)~Experimental setup of two capacitively coupled ($C_0$) RLC circuits with individual tuning voltages $U_{1,2}$. (b)~Response $v_{1}$ and $v_{2}$ of the two resonators in a sweep from low to high frequency $f_d$. 
  The system switches between the 0-state ($v_{1,2}=0$), S-state [$\text{sign}(v_1)=\text{sign}(v_2)$], A-state [$\text{sign}(v_1)=-\text{sign}(v_2)$], or M-state. (c)~Diagram characterizing the stable solutions measured in frequency upsweeps. Blue and red intensity mark the absolute value of the symmetric and antisymmetric combination of $v_{1,2}$, $\left|v_{\rm S} \right| = \left|v_1 + v_2\right|/\sqrt{2}$ and $\left|v_{\rm A} \right| = \left|v_1 - v_2\right|/\sqrt{2}$, respectively (analogous graphs can be drawn for $u_{S,A}$). The resulting diagram shows regions in white, red, blue, and purple, corresponding to measured 0, S, A, and M-states, respectively. The brown line is calculated from theory, see Eq.~\eqref{eq:brownline}. (d)~In total there are up to 9 stable solutions at any point in the phase diagram, which can be characterized by their symmetry. (e)~and (f)~Schematic representation of the steady-state solutions and bifurcation points. Solid lines: stable; dashed: singly unstable, i.e., one characteristic exponent has a positive real part~\cite{Soriente_2021}; dotted: doubly unstable, i.e., two characteristic exponents have positive real parts. Squares (circles) indicate bifurcations that involve only unstable  (stable and unstable) solutions. Their colors match the lines from (g).  (g)~Calculated stability phase diagram. White: 0-state is stable; blue: only the S-state is stable; red: S and A-states are stable; purple: S and M-states are stable; dark red: S, A and M-states are stable. Circle and squares represent bifurcations from (e) and (f). }
  \label{fig:fig2}
\end{figure*}

\textit{Experimental setup -- } We built a setup of capacitively coupled parametrons using electrical lumped-element resonators, see Fig.~\ref{fig:fig2}(a). Each resonator (marked by index $i$) comprises a resistance $R = \SI{47}{\mega\ohm}$, an inductance $L = \SI{87}{\micro\henry}$, a tuning voltage $U_i \approx \SI{2}{\volt}$, and a nonlinear capacitance $C\approx\SI{20}{\pico\farad}$ in the form of a varicap diode. The resonators are driven and read out inductively through auxiliary coils. Using Kirchhoff's laws, our electrical circuits are described by coupled equations of motion~\cite{Nosan_2019},
\begin{multline}
	\ddot{x}_i + \omega_i^2\left[1-\lambda\cos\left(2\omega_d t\right)\right]x_i + \alpha_i x_i^3 \\+ \gamma_i \dot{x}_i -\sum_{j\neq i}J_{ij} x_j = 0\,. \label{eq:coupled_EOM}
\end{multline}
Here, dots indicate time derivatives, $x_i = u_i \cos(\omega t) - v_i \sin(\omega t)$ is the measured voltage with quadrature amplitudes $u_i$ and $v_i$, $\omega_i = 2\pi f_i$ is the angular resonance frequency, $\alpha_i$ the coefficient of the Duffing nonlinearity, $\gamma_i = \omega_i/Q_i$ the damping rate, and  $Q_i$ the quality factor of the $i^{\rm th}$ resonator.
Our resonators are (nearly) identical in their bare characteristics and are tuned via $U_i$ to have (nearly) degenerate eigenfrequencies, $\omega_i \approx \omega_0 = 2\pi f_0$. They are linearly coupled with coefficients $J_{ij}$ ($i\neq j$) and are all driven with the same parametric modulation depth $\lambda = 2U_d/(U_{th}Q)$ at an angular rate $2\omega_d = 4\pi f_d \approx 2\omega_0$, where $U_d$ is a driving voltage and $U_{th}$ is the corresponding threshold voltage for parametric oscillations. For further details on the individual resonators, cf. Ref.~\cite{Nosan_2019}.

\textit{Slow-flow -- } We measure the system with a lock-in amplifier, and are thus primarily interested in changes of the slow coordinates $(u_i, v_i)$ on timescales much longer than $1/\omega_0$. We compare the observed results with calculated stationary states of a slow-flow treatment~\cite{guckenheimer_1990, Papariello_2016, Heugel_2019_TC, Soriente_2021} of our model~\eqref{eq:coupled_EOM}, which is equivalent to a rotating mean-field analysis of the corresponding quantum system. Specifically, we obtain coupled equations of motion for the `slow' order parameters
\begin{align}
    \label{eq:slowflow}
    \dot{u}_i &= -\frac{\gamma  u_i}{2}- \left(\frac{3 \alpha}{8 \omega_d }X_i^2 +\frac{\omega_0^2 - \omega_d^2}{2\omega_d}+\frac{\lambda \omega_0^2}{4 \omega_d }\right) v_i +\frac{J  v_j}{2 \omega_d }\,,\nonumber\\
    \dot{v}_i &= -\frac{\gamma  v_i}{2} + \left(\frac{3 \alpha}{8 \omega_d }X_i^2 + \frac{\omega_0^2 - \omega_d^2 }{2\omega_d} -\frac{\lambda \omega_0^2}{4 \omega_d }\right)u_i-\frac{J  u_j}{2 \omega_d }\,,
\end{align}
where $X_i^2 = u_i^2 + v_i^2$. Our method is valid since $\lambda$, $\gamma/\omega_0$, $J/\omega_0^2$ and $(\alpha/\omega_0^2)x_i^2$ are all of order $\epsilon$ with $0<\epsilon\ll 1$~\cite{nayfeh2008}. The stationary states are obtained by solving the polynomial equations~\eqref{eq:slowflow} for $\dot{u}_i=\dot{v}_i=0$. Note that we assumed here homogeneous dissipation $\gamma_i = \gamma$, nonlinearities $\alpha_i = \alpha$, and coupling $J_{ij}=J$. Since the nonlinearity $\alpha$ can be eliminated by rescaling $u_i$ and $v_i$ with $\sqrt{\alpha}$, the magnitude of $\alpha$ only changes the amplitudes of $u_i$ and $v_i$.
In coupled systems, working in the normal mode basis can be very helpful. The symmetric and antisymmetric eigenmode are shifted to $\sqrt{\omega_0^2\pm J}$ and parametric excitation around these frequencies are expected. As the Duffing nonlinearity mixes the normal modes, several nonlinear coupling terms appear (see Appendix~\ref{sec:nonlinear_coupling}), and for brevity's sake, we remain in the bare mode notation.

\textit{Measurement protocol -- } Following standard  protocols~\cite{Leuch_2016,Heugel_2019_TC,Nosan_2019}, we characterize the independent circuits (for vanishing $J$) by sweeping $f_d$ for different $U_d$, see Appendix~\ref{sec:single_prm}.
We then couple two such devices capacitively and perform similar sweeps with increasing $f_d$ (upsweep) to probe the system's stationary states, see Fig.~\ref{fig:fig2}(b).
For $U_d = \SI{2.5}{\volt}$, the sweep yields two frequency segments with large responses. In one of them, the two resonator oscillations are in phase and virtually identical (`symmetric' = S). In the other segment, the oscillations are out of phase; above \SI{2.36}{\mega\hertz}, the amplitudes are approximately identical (`antisymmetric' = A), while below \SI{2.36}{\mega\hertz}, the amplitudes differ significantly (`mixed-symmetry' = M). When increasing the parametric drive to $U_d = \SI{3.7}{\volt}$, the response segments merge and the system directly jumps from the A-state to the S-state response slightly above \SI{2.36}{\mega\hertz}.

\textit{Measured phase diagram -- } To obtain the full measured stability diagram of the two-parametron system, we repeat the frequency sweeps as in Fig.~\ref{fig:fig2}(b) for a wide range of  driving voltages $U_d$, see Fig.~\ref{fig:fig2}(c).
The overall shape is that of two normal modes with partially overlapping instability lobes. From the mode splitting and the observation that the antisymmetric mode appears at a lower frequency than the symmetric mode, we extract $J = \SI{-1.28}{\mega \hertz^2}$. Surprisingly, however, the transition from A-~to S-oscillation states does not occur at the boundaries of the individual lobes; rather, the S-region protrudes to the left at $U_d\approx\SI{3.5}{\volt}$, and then proceeds along a diagonal line. In addition, we observe that the M-state appears first along the upsweep in some parts of  the lower left lobe. 
Crucially, the M-state and the reduced region of A-oscillations are unexpected from a naive weak coupling perspective~\cite{Rota_2019, Goto_2016, Puri_2017_NC, Dykman_2018, Heugel_2019_TC}. As we discuss below, they arise due to an interplay between the coupling and the negative nonlinearity.

\textit{Stability analysis -- } To better understand the measured phase diagram and the validity of different regions for Ising computation, we solve for the stationary stable states of Eqs.~\eqref{eq:slowflow}. According to Bézout’s theorem, four cubic equations have maximally $3^4$ roots. For the case of a single oscillator it has been shown that the relevant solution space comprises three stable and two unstable states~\cite{Lifshitz_Cross,Papariello_2016}. Consequently, in the limit of weak coupling, we expect for two-oscillators up to 25 physical solutions, with maximally nine stable physical states, depending on the drive amplitude and detuning~\cite{stability}. Although we consider coupling strengths beyond that limit, we did not observe any additional steady state solutions. Furthermore, in our time-dependent numerical analysis and the experiment, we did not see any periodic states in the rotating frame, which might appear for such coupled systems~\cite{Strinanti_2021}.  

In Fig~\ref{fig:fig2}(d), the state at the origin of the coordinate system (0) indicates that all resonators are at zero amplitude. The S and A-states correspond to the parametron phase states of the system's normal mode solutions, where the two resonators oscillate at the same amplitude either in phase or with opposite phase. These solutions can, in principle, be used for Ising simulation and be interpreted as ferromagnetic and antiferromagnetic spin configurations.
The other four solutions have no simple symmetry and are mixed-symmetry states (M).

\textit{Calculated phase diagram -- } We map the system's phase diagram by tracking the stability of the the different oscillation states in the parameter space spanned by $f_d$ and $U_d$. We use in our analysis the experimentally determined parameters for $f_0$ and $J$, while $Q=265$ was chosen for both resonators to achieve optimal agreement. In our system, all transitions are of $Z_2$ spontaneous symmetry-breaking type (pitchfork bifurcations)~\cite{Soriente_2021}. For consistency with literature in the field, we refer to them as phase transitions~\cite{Soriente_2021}. For $U_d = \SI{2.5}{\volt}$ [Fig.~\ref{fig:fig2}(e)], from high to low frequencies, the system transitions from zero amplitude (0-state), to parametron S-states, to coexistence of the 0 and S-states, to coexistence of S and A-states, to coexistence of M and S-states, and finally to coexistence of M, S, and 0-states. Correspondingly, we identify that along the experimental upsweep in the left panel of Fig.~\ref{fig:fig2}(b), the system jumps along the states 0-M-A-0-S-0. For $U_d = \SI{3.7}{\volt}$ [Fig.~\ref{fig:fig2}(f)], from high to low frequencies, the system transitions from the 0-state, to S-states, to coexistence of the S- and A-states, to coexistence of M, S, and A-states, and finally to coexistence of all nine solutions. In the experimental upsweep in the right panel of Fig.~\ref{fig:fig2}(c), the system jumps along the states 0-A-S-0. Based on this methodology, we find which states appear in a measured phase diagram as a function of $f_d$ and $U_d$, compare Figs.~\ref{fig:fig2}(g) and (c).

The most prominent feature in both Figs.~\ref{fig:fig2}(c) and (g) is the phase transition at the solid brown line. This transition arises from an interplay between the nonlinearity and the inter-parametron coupling which, induces a parametric coupling of the S and A-states, cf.~Eq.~\eqref{eq:coupled_EOM}, see Appendix~\ref{sec:nonlinear_coupling}. In an upsweep, the brown line marks the gradual transition from M- to A-oscillations (below $U_d\approx\SI{3.5}{\volt}$) or the sharp jump from A- to S-states (above $U_d\approx\SI{3.5}{\volt}$). Its position $\lambda_A(\omega)$ is obtained by a stability analysis against small fluctuations~\cite{Soriente_2021}), which yields the expression
\begin{align}
\label{eq:brownline}
\lambda_A = \frac{2\sqrt{ \gamma ^2 \omega ^2+\left(2J- \left(\omega ^2-\omega_0^2\right)\right)^2}}{\omega_0^2}\,.
\end{align}
Note that only the sign of $\alpha$ impacts the position of $\lambda_A$, but not its magnitude as $\alpha$ only rescales $u_i$ and $v_i$. Consequently, as the coupling coefficient $J$ is decreased (increased), the stability boundary of the A-state  approaches (recedes from) the right boundary of the antisymmetric instability lobe. This effect bears important implications when using parametron networks for Ising machines, as it reduces the regions of the phase diagram where the A-oscillation states are stable.

\section{Discussion}\label{sec:discussion}

Our network of two coupled parametrons realizes the smallest classical form of an `adiabatic bifurcation' simulator~\cite{Goto_2016,Nigg_2017,Puri_2017_NC}, but in the unexplored regime of strong coupling. In general, the standard operation protocol for using such a simulator is as follows: first, initialize the network in its 0-state with $\lambda = 0$ and $\Delta < -\lvert J \rvert/\omega_0$, where $\Delta = \omega_d-\omega_0$ is the detuning. Second, increase the parametric drive $U_d \propto \lambda$ to push all of the individual resonators across their parametric threshold $\lambda_\mathrm{th}$, such that they all ring up to a phase state. Due to the mutual coupling between the resonators, certain phase state configurations (symmetries) will be favored over others. Numerical simulations for the case of two parametrons predict that this favored oscillation mode is the ferromagnetic (S) state for $J>0$ and the antiferromagnetic (A) state for $J<0$~\cite{Nigg_2017, Puri_2017_NC}. Presumably, a measurement of the final oscillation state can therefore be used to find the ground state of the corresponding Ising Hamiltonian
\begin{align}\label{eq:Ising_H}
    H_\mathrm{Ising} = -\sum_{i,j} J_{i,j}\sigma_i\sigma_j\,,
\end{align}
where $\sigma_i \in \{-1,1\}$ is the classical state of a spin that points either up or down. Measuring the S-state of our two-parametron network at the end of the protocol is then associated with a ground state $\sigma_1 = \sigma_2$, while the A-state corresponds to $\sigma_1 = -\sigma_2$. Other protocols, by contrast, rely on $H_\mathrm{Ising}$ as an effective description of the parametron network~\cite{Dykman_2018}.

\begin{figure}[t!]
  \includegraphics[width=\columnwidth]{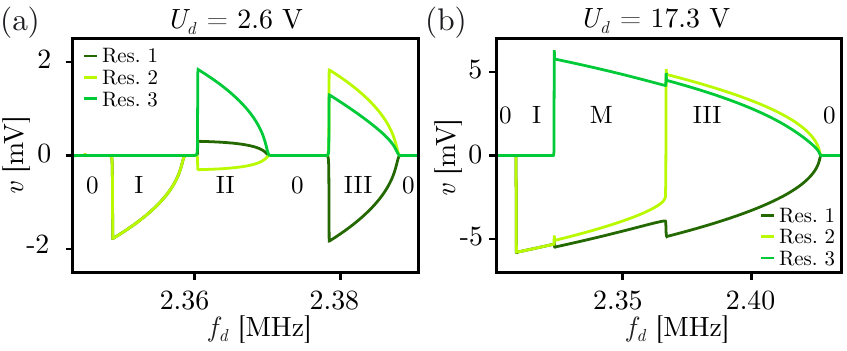}
  \caption{(a)~Numerical simulations of three identical resonators with the same parameters as our experimental devices, and with coupling coefficients $J_{1,2} = -2 J$, $J_{1,3} = -0.6 J$ and $J_{2,3} = 0.6 J$. When sweeping the parametric drive from low to high frequencies at $\lambda = 1.5\lambda_\mathrm{th}$, we observe the instability lobes I to III of the network normal modes. The three-parametron phase states of II and III can be interpreted as the Ising states $\pm(1,-1,1)$ and $\pm(-1,1,1)$, respectively. Within lobe I, resonator 3 has amplitude 0, which makes it difficult to map this three-parametron state to a particular spin configuration. (b)~For $\lambda = 10\lambda_\mathrm{th}$, between states I and II we find a mixed-symmetry state (M) that reflects the  ground state of the corresponding Ising Hamiltonian Eq.~\eqref{eq:Ising_H}.}
  \label{fig:fig3}
\end{figure}

From our results, we reveal crucial additional understanding of adiabatic bifurcation protocols and their validity range for Ising machines. Effectively, our upsweep scans in Figs.~\ref{fig:fig2}(b)-(c) follow exactly a bifurcation simulator operation protocol, where (without loss of generality) we chose to sweep $f_d$ at constant $U_d$, rather than vice versa. We start in the 0-state and evolve into the instability regions to obtain, at sufficiently high $U_d$, the correct outcome of the computation, i.e., the antiferromagnetic (A) state. At the same time, our result highlights key important differences to what is commonly discussed in the literature: on the one hand, we found up to nine stationary oscillation configurations, cf. the M-states in Figs.~\ref{fig:fig2}(e)-(g); these go beyond the solution space of the Ising mapping. Extrapolating our discussion on two parametrons, we expect that a network of $N$ parametrons can form up to $3^N$ stable states which the system will explore in certain parameter regimes. On the other hand, in the operation protocol presented above, it turns out that the outcome of the computation relies on `which mode has the lower eigenfrequency'. The problem is thus mapped to the linear splitting of the normal modes, which does not scale with $2^N$ but with $2N$ ($N$ normal modes with two phase states each). Our network has the same solution space as expected from an Ising system only by coincidence because $2^N=2\times N$ for $N=2$.

\begin{figure}[t!]
  \includegraphics[width=\columnwidth]{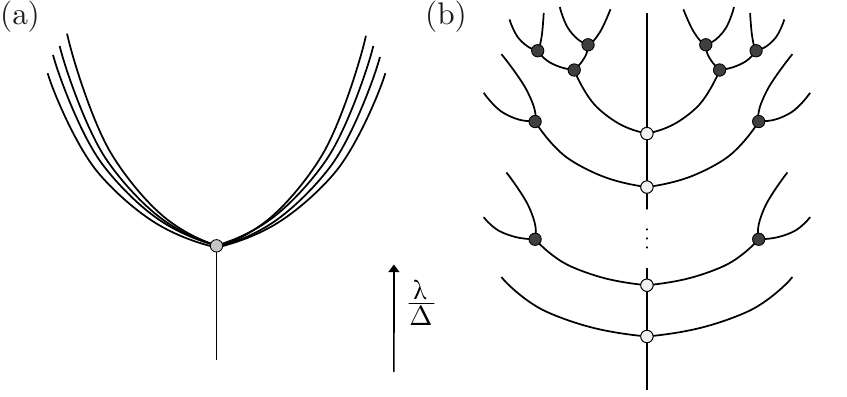}
  \caption{(a)~Schematic illustration of $N$ (almost) identical uncoupled, parametrically driven resonators. When sweeping the driving strength $\lambda$ versus the detuning $\Delta$, each resonator undergoes a phase transition into parametron phase states at the same threshold, see grey dot. Above the threshold, the system is described by the combinations of $N$ bistable states, offering $2^N$ possible many-body states. (b)~Schematic illustration of $N$ (almost) identical coupled, parametrically driven resonators. The coupling leads to the formation of $N$ normal modes involving all resonators. Each mode undergoes a phase transitions to parametron phase states at an individual threshold value during the sweep, see white dots. Each of the $2N$ possible normal-mode phase states then branches into mixed-symmetry states due to the interaction with other normal modes, see black dots. In total, it is expected that the system will feature up to $2^N$ states far above the last threshold, corresponding to the case of negligibly small coupling relative to the influence of the drive. In this simplified view, we have neglected details such as saddle-node bifurcations, and we do not explicitly differentiate between stable and unstable solutions of the system. An example of this qualitative picture manifests for $N=2$ in Fig.~\ref{fig:fig2}(f).}
  \label{fig:fig4}
\end{figure}

The discrepancy between the normal-mode picture and the Ising mapping becomes more acute when generalizing the above recipe to networks with $N>2$ resonators. As we have established, the oscillation states and the corresponding instability lobes are inherited from the normal modes of the underlying coupled resonator system, of which there are precisely $N$. The number of available many-body parametron phase states is therefore $2N$. The number of Ising configurations, on the other hand, must be $2^N$, which is larger than $2N$ for any $N>2$. From this argument, it follows directly that there can be no one-to-one mapping between the normal modes and the possible Ising configurations. Indeed, as we show with a numerical example in Fig.~\ref{fig:fig3}(a), the stable solutions of the network can deviate from the naively expected combination of resonator phase states with equal amplitudes. 
We can therefore not trust the ordering of the normal modes in frequency to find the ground state solution of the Ising problem.

How can one still reconcile our observations with the predicted operation of adiabatic bifurcation Ising machines~\cite{Goto_2016, Nigg_2017, Puri_2017_NC, Dykman_2018}? Far beyond the parametric threshold, it is generally assumed that each parametron occupies one of its individual (uncoupled) phase states, giving rise to $2^N$ possible oscillation states, see the simulation results in Fig.~\ref{fig:fig3}(b) and the schematic illustration in Fig.~\ref{fig:fig4}(a)~\cite{Goto_2016, Nigg_2017, Puri_2017_NC, Dykman_2018}.
It is at present not clear, however, how the transition from a normal-mode regime with $2N$ states ($\lambda \gtrsim \lambda_\mathrm{th}$) to an Ising regime with $2^N$ states ($\lambda \gg \lambda_\mathrm{th}$) takes place in detail. We tentatively propose that this transition involves mixed-symmetry states similar to the ones we found in our system -- as an illustration, the simulation with $N=3$ in Fig.~\ref{fig:fig3}(b) clearly shows mixed-symmetry states that approximate Ising configurations for $\lambda = 10\lambda_\mathrm{th}$. As a result, the simulated network has $2^N$ solutions in the center region between roughly \SI{2.325}{\mega\hertz} to \SI{2.375}{\mega\hertz}. Furthermore, mixed-symmetry states can even increase the solution space up to $3^N$, which is $>2^N$ and therefore \textit{beyond} the scope of the Ising model. Indeed, in the simulation, additional I-states appear below \SI{2.325}{\mega\hertz} that have no analog in an Ising spin system. In Fig.~\ref{fig:fig4}(b), we schematically sketch how cascades of bifurcations, branching off from the $N$ normal mode states in the central `trunk' line, can form an Ising network with $2^N$ states for a hypothetical optimal trajectory through the phase diagram.

We emphasize that the discrepancy between the normal-mode regime and the Ising regime should occur in every system of coupled parametrons, without regard to the coupling strength and coupling type (dissipative or bilinear). However, the region of the phase diagram over which the transition occurs should depend on the relative strength of $J$, $\alpha$ and $\lambda$. For very small coupling, this transition could take place close to the threshold, such that it may be overlooked. Note that it was recently observed that Coherent Ising Machines have a higher probability of finding the correct solution for large driving than directly above the threshold~\cite{Strinanti_2021}. This could potentially be explained by the transition discussed here.

The system we present in this work is entirely classical. Here, one particular state is chosen at every phase transition, depending on the instantaneous boundary condition (e.g. thermal fluctuations prior to or during the transition). In the presence of sufficiently small noise, this solution is then stable over long timescales. Our discussion regarding the solution space of an Ising machine, however, is equally relevant for quantum systems with strong Kerr nonlinearities~\cite{Grimm_2019}. There, the mean photon number of each resonator is small, and each phase transition leads to a quantum superposition of the branching coherent states~\cite{Goto_2016,Nigg_2017,Puri_2017_NC,Puri_2019_PRX}. Using this superposition to find the ground state of an Ising problem (with $2^N$ spin configurations) is only possible if the right solution space is available. However, as we show above, the solution space can vary between $2N$ states (which is too small a number for any $N>2$) up to $3^N$ states (which is too large). Understanding in detail the phase diagram of a system of coupled parametrons is therefore of fundamental importance for quantum adiabatic Ising machines.

\section{Outlook} 

As coupled networks of parametric resonators are one of the main candidates for future parallel computation architectures, our study provides crucial input for a growing community working towards classical and quantum analog computation~\cite{Gottesman_2001, Devoret_2013, Mahboob_2016, Inagaki_2016, Goto_2016,Nigg_2017,Dykman_2018,Puri_2019_PRX}. Furthermore, it provides additional incentive for the fundamental exploration of complex driven-dissipative nonlinear networks in a multitude of fields~\cite{DykmanBook}. Future experimental and theoretical research will address the transition between regimes with $2N$, $2^N$ and $3^N$ states, and provide concrete recipes how the correct solution space can be selected. Additionally, the discussion will be extended to classical and quantum `Boltzmann machines' that operate with fluctuations to anneal the network into an optimal many-body state~\cite{Goto_2018}.

\acknowledgments
This work received financial support from the Swiss National Science Foundation through grants (CRSII5\_177198/1) and (PP00P2\_163818). We thank Peter M\"{a}rki and \v{Z}iga Nosan for technical help. We thank A. Grimm, E. Dalla Torre and M. C. Strinati for stimulating discussions.


\appendix

\section{Single parametron characterization}
\label{sec:single_prm}
\begin{figure}[t!]
  \includegraphics[width=\columnwidth]{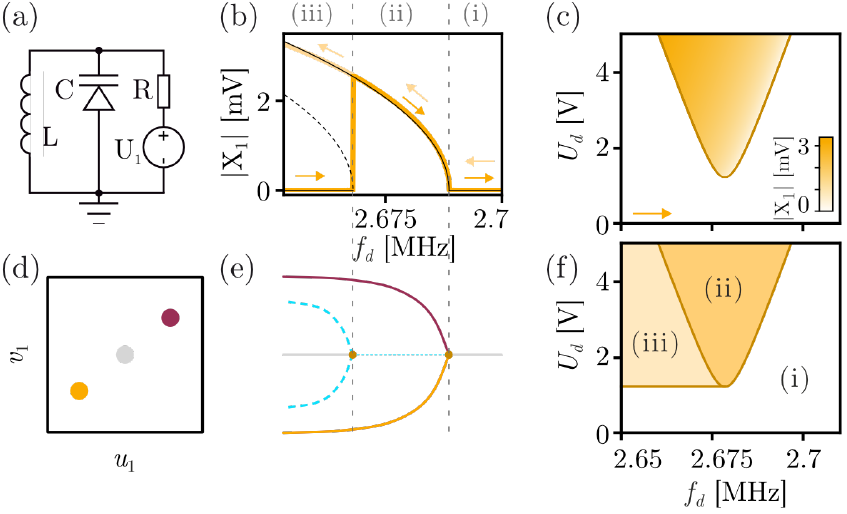}
  \caption{Characterization of a single parametron. (a)~Schematic of a parametron realization in the form of an RLC circuit with resistance $R$, inductance $L$, nonlinear capacitance $C$, and tuning voltage $U_{1}$. (b)~Amplitude response $X_1$ (orange lines) of the parametric resonator measured with a lock-in amplifier as a function of frequency $f_d$ at a driving strength of $U_d=\SI{3}{\volt}$. Arrows indicate the sweep direction. Thin gray solid (dashed) lines are calculated stable (unstable) steady-state solutions using the rotating steady-states of Eq.~\eqref{eq:coupled_EOM} [cf.~Eqs.~\eqref{eq:slowflow}] with $\lambda_1=\frac{U_d}{U_{th}}\frac{2}{Q}=0.017$, and  $\alpha_1 = \SI{-9e17}{\per\square\volt\per\square\second}$ used as a fit parameter~\cite{Leuch_2016}. (c)~$X_1$ measured as a function of $U_d$ and $f_d$ using upsweeps (arrow) at constant $U_d$, showing a typical parametric instability lobe. 
  The solid line is a theory fit to the lobe boundaries that allows for estimation of $Q = 295$. (d)~Schematic representation in phase space of the parametron phase states (wine red, orange), and the 0-amplitude state (grey). (e)~Schematic representation of the steady-state solutions and bifurcation points as a function of $f_d$. Solid (dashed) lines indicate stable (unstable) solutions. (f)~Calculated stability phase diagram. (i) White: only the 0-amplitude solution is stable; (ii) orange: only the phase states are stable; (iii) light orange: 0-amplitude and phase states are both stable. The regions (i)-(iii) manifest in (b) and (e).
  }
  \label{fig:FigA1}
\end{figure}

We characterize the independent circuits and recover their characteristic parametric instability lobes following standard characterization protocols~\cite{Leuch_2016,Heugel_2019_TC}. When decoupled from one another (vanishing $J$), each resonator can be driven into parametric resonance when $U_d \geq U_{th}$~\cite{Landau_Lifshitz, Lifshitz_Cross}.
In Fig.~\ref{fig:FigA1}(b), we show experimental sweeps with increasing and decreasing $f_d$ for constant $U_d$, exhibiting the standard nonlinear parametric response and hysteresis of the parametron labelled as 1 (similar results were obtained for parametron 2). Inside the region marked as (ii), the linear resonator becomes unstable, bifurcates and settles into one of the two phase states that are stabilized by $\alpha$~\cite{Lifshitz_Cross}. In region (iii), the phase states coexist with a stable solution at $X_1=0$. Repeating the upsweeps (increasing $f_d$) for different $U_d$, we recover the characteristic parametric instability lobe, cf. Fig.~\ref{fig:FigA1}(c).

Using Eqs.~\eqref{eq:slowflow}, we can describe all of the measured results of the single parametron~\cite{Papariello_2016, Leuch_2016}, by comparing to the calculated steady-state amplitudes $X_1$. From a mathematical point of view, our stationary order parameters are obtained by solving a quintic characteristic polynomial. We obtain up to three different stable states (attractors) in phase space, cf. Fig.~\ref{fig:FigA1}(d). As a function of $f_d$, phase transitions occur as the number of stable solutions changes at specific `bifurcation points'. 

In the single decoupled parametrons, we observe a second-order continuous (first-order discontinuous) time-translation $Z_2$ symmetry-breaking phase transition, a.k.a.~period-doubling bifurcation, when the zero amplitude mode continuously splits into the two phase state at the (i)--(ii) [(ii)--(iii)] boundary, cf. Fig.~\ref{fig:FigA1}(e). Fitting the model to the measurement data, we determine the values $Q_1 = 295$, $f_0 = \SI{2.6784}{\mega\hertz}$, $\alpha_1 = \SI{-9e17}{\per\square\volt\per\square\second}$, and $U_{th} = \SI{1.21}{\volt}$~\cite{Leuch_2016, Nosan_2019}. In particular, from the fact that region (iii) appears at $f_d < f_0$, we infer that $\alpha<0$~\cite{Eichler_2018}. 
The corresponding parameter characterization of the coupled system studied in the main text yielded $Q = 265$, $f_0 = \SI{2.3670}{\mega\hertz}$, $\alpha = \SI{-6.5e17}{\per\square\volt\per\square\second}$, and $U_{th} = \SI{1.73}{\volt}$.

\section{Nonlinear coupling between normal modes}\label{sec:nonlinear_coupling}

A system of two, nearly identical, coupled linear resonators can be described in terms of uncoupled normal modes by moving to symmetric and antisymmetric coordinates. Nonlinearities will then generally couple the symmetric and antisymmetric modes.
For our nonlinear parametric system, the slow-flow equations for ($u_S$, $v_S$) are given by:
\begin{widetext}
\begin{equation}
\begin{pmatrix}
\dot{u}_S\\
\dot{v}_S
\end{pmatrix}=
\begin{pmatrix}
     -\frac{v_S \left(-J+\frac{3}{4} \alpha  \left(u_A^2+v_A^2\right)-\omega_d ^2+\omega_0^2\right)}{2 \omega_d }-\frac{v_S \left(4 \lambda  \omega_0^2-3 \alpha  \left(u_A^2-v_A^2\right)\right)}{16 \omega_d }-\frac{ 3 \alpha  u_A u_S v_A }{8 \omega_d }-\frac{3 \alpha  v_S \left(u_s^2+v_S^2\right)}{16 \omega_d }-\frac{\gamma u_S }{2}\\
    \frac{u_S \left(-J+\frac{3}{4} \alpha  \left(u_A^2+v_A^2\right)-\omega_d ^2+\omega_0^2\right)}{2 \omega_d }-\frac{u_S \left(4 \lambda  \omega_0^2-3 \alpha  \left(u_A^2-v_A^2\right)\right)}{16 \omega_d }+\frac{3 \alpha  u_A v_A v_S}{8 \omega_d }+\frac{3 \alpha  u_S \left(u_S^2+v_S^2\right)}{16 \omega_d }-\frac{\gamma v_S }{2}
\end{pmatrix}\,.
\end{equation}
\end{widetext}
Note that because of the Duffing nonlinearity, the effective eigenfrequency of the symmetric mode, $\omega_0^2-J \rightarrow \omega_0^2-J +\frac{3}{4} \alpha  \left(u_A^2+v_A^2\right)$, as well as its parametric driving strength, $4\lambda  \omega_0^2 \rightarrow 4\lambda  \omega_0^2-3 \alpha  \left(u_A^2-v_A^2\right)$ now explicitly depends on the coordinates ($u_A$, $v_A$) of the antisymmetric mode change. In addition to the usual expected bifurcations for both normal modes, the aforementioned interplay triggers a further bifurcation, cf. brown point in Fig.~\ref{fig:fig2}(e), which heralds the mixed-symmetry state (M). At this bifurcation, the oscillations of the antisymmetric mode are strong enough to drive parametric oscillations of the symmetric mode through the nonlinearity, leading to the mixed-symmetry state. This effect takes place in the instability lobe with lower (higher) eigenfrequency for negative (positive) $\alpha$.


%

\end{document}